\documentclass[aps,pra,twocolumn,showpacs,groupedaddress,floatfix]{revtex4}
\usepackage{graphicx}
\usepackage[latin1]{inputenc}
\usepackage{amsmath}
\usepackage{amssymb}
\usepackage{color}

\newcommand{\nc}{\newcommand}
\nc{\bra}[1]{\langle #1|}
\nc{\ket}[1]{|#1\rangle}
\nc{\braket}[1]{\left\langle #1 \right\rangle}
\nc{\equ}[1]{\begin{eqnarray*}#1\end{eqnarray*}}
\nc{\equn}[1]{\begin{eqnarray}#1\end{eqnarray}}
\nc{\dagg}{^{\dagger}}
\nc{\conj}{^{*}}
\nc{\dx}[1]{\, \mathrm{d} {#1} \,}
\nc{\Dx}[1]{\mathcal{D} {#1} \,}
\nc{\la}{\langle}
\nc{\ra}{\rangle}
\nc{\tr}{\text{tr}}
\nc{\Tr}{\text{Tr} \,}
\nc{\e}{\text{e}}
\nc{\Id}{\mathbb{1}}
\nc{\eps}{\varepsilon}
\nc{\der}[2]{\frac{\mathrm{d} {#1}}{\mathrm{d} {#2}}}
\nc{\pder}[2]{\frac{\partial {#1}}{\partial {#2}}}
\nc{\bigO}{\mathcal{O}}
\nc{\half}{\frac{1}{2}}

\nc{\Eq}[1]{Eq.~(\ref{#1})}
\nc{\eq}[1]{Eq.~(\ref{#1})}
\nc{\chap}[1]{Chapter \ref{#1}}
\nc{\Sect}[1]{Section \ref{#1}}
\nc{\sect}[1]{section \ref{#1}}
\nc{\fig}[1]{Fig.~\ref{#1}}
\nc{\Fig}[1]{Fig.~\ref{#1}}
\nc{\tabl}[1]{Table \ref{#1}}
\nc{\app}[1]{Appendix \ref{#1}}

\nc{\eg}{\emph{e.g.} }
\nc{\ie}{\emph{i.e.} }
\nc{\etal}{et al.}

\nc{\asp}[1]{\textcolor{red}{[{\bf SP}: #1]}}
\nc{\alg}[1]{\textcolor{blue}{[{\bf ALG}: #1]}}

\begin{document}

\title{Cavity QED simulation of qubit-oscillator dynamics in the ultrastrong coupling regime}

\author{Arne L. Grimsmo}\email{arne.grimsmo@ntnu.no}
\affiliation{Department of Physics, 
             The Norwegian University of Science and Technology, N-7491 Trondheim, Norway}
\affiliation{Department of Physics, 
             University of Auckland, Private Bag 92019, Auckland, New Zealand}
\author{Scott Parkins}\email{s.parkins@auckland.ac.nz}
\affiliation{Department of Physics, 
             University of Auckland, Private Bag 92019, Auckland, New Zealand}
\date{\today}

\begin{abstract}
We propose a quantum simulation of a two-level atom coupled to a single mode of the electromagnetic field in the ultrastrong coupling regime based upon resonant Raman transitions in an atom interacting with a high finesse optical cavity mode. We show by numerical simulation the possibility of realizing the scheme with a single rubidium atom, in which two hyperfine ground states make up the effective two-level system, and for cavity QED parameters that should be achievable with, for example, microtoroidal whispering-gallery-mode resonators. Our system also enables simulation of a generalized model in which a nonlinear coupling between the atomic inversion and the cavity photon number occurs on an equal footing with the (ultrastrong) dipole coupling and can give rise to critical-type behavior even at the single-atom level. Our model takes account of dissipation, and we pay particular attention to observables that would be readily observable in the output from the system.
\end{abstract}

% 42.50.Fx?
%\pacs{42.50.Pq,05.70.Fh,03.67.Ac,42.50.Ct,32.80.-t,42.60.Da}
\maketitle

\section{Introduction}

The idea of analog quantum simulations, that is, engineering one quantum system to replicate the behaviour of another that one wishes to study, was first proposed by Feynman \cite{Feynman82} with regards to the infeasibility of simulating quantum systems on classical computers. The simulating system typically involves a larger number of degrees of freedom, and the effective simulation relies on precise variation of system parameters through exquisite experimental control. This idea has become reality in recent years \cite{Buluta09,Cirac12} and fundamental models of interacting quantum systems have been realized thanks, for example, to advances in the control and manipulation of ultracold gases in optical lattices \cite{Bloch08}. Systems based on cavity quantum electrodynamics (cavity QED) also offer exciting possibilities, as exemplified by the recent demonstration of the Dicke quantum phase transition with a superfluid atomic gas in an optical cavity \cite{Baumann10}, in which a pair of discrete momentum states of the gas was used to simulate the two-level atoms of the original Dicke model of superradiance \cite{Dicke54,Hepp73}.

In this work we focus on a closely related model -- the Rabi model -- which describes the interaction of a two-state system (atom or qubit) with a single quantized harmonic oscillator (electromagnetic field mode) through the Hamiltonian
\begin{align}\label{eq:Rabi}
  H_R = \frac{\omega_0}{2} \sigma_z + \omega a\dagg a + g(\sigma_+ + \sigma_-)(a+a\dagg) ,
\end{align}
where $\omega_0$ and $\omega$ are the qubit and oscillator frequencies, respectively, $g$ is the interaction strength, $\{\sigma_z,\sigma_\pm\}$ are two-state operators, and $a$ ($a\dagg$) is the annihilation (creation) operator for the oscillator.
This Hamiltonian predicts accurately many physical situations where an atom -- artificial or real -- is interacting with a confined cavity field. However, when this model is used in the description of conventional cavity QED systems, it is generally the case that the coupling constant $g$ is much (i.e., orders of magnitude) smaller than the frequencies $\omega_0$ and $\omega$. This means that terms in $H_R$ that do not conserve the total excitation number can be neglected, in what is known as the ``rotating wave approximation.'' This leads to probably the most studied model in quantum optics, the Jaynes-Cummings model \cite{Jaynes63}:
\begin{align}\label{eq:JC}
  H_{JC} = \frac{\omega_0}{2} \sigma_z + \omega a\dagg a + g(\sigma_+ a + \sigma_- a\dagg).
\end{align}
Recently, however, going beyond this approximation has gained newfound interest, as novel experimental systems have pushed the boundaries for the strength of the coupling between (artificial) atoms and cavity field modes. In particular, in experiments using circuit QED \cite{Schoelkopf08} and semiconductor microcavities, couplings between artificial atoms and cavity modes have reached the so-called ``ultrastrong'' regime \cite{Gunter09,Anappara09,Niemczyk10,Forn-Diaz10}. Here, values of $g/\omega \sim 10\%$ have been realized, so that $g$ is no longer small compared to the mode frequency. 

These experimental advances have in turn stimulated new theoretical investigations (see, for example, \cite{Irish05,Liu09,Pan10,Hausinger10,Braak11,Chen12}), as the conventional treatment is no longer adequate. To appreciate this necessity, notice that for the Jaynes-Cummings model, $H_{JC}$ conserves the total excitation number (i.e., $\frac{1}{2}(\sigma_z+1) + a\dagg a$), and, together with conservation of energy, this means that the spectrum is easily found. If, on the other hand, the terms $a\sigma_-$ and $a\dagg \sigma_+$ cannot be ignored, then this is no longer the case, and until recently no analytical solution was known. This was remedied in \cite{Braak11} by Braak, who showed that a discrete symmetry, the conservation of parity $\Pi = -\sigma_z(-1)^{a\dagg a}$, is enough for the model to be integrable and derived analytical expressions for the spectrum (see also \cite{Chen12}). Even though some intuition can be drawn from the conservation of parity \cite{Casanova10,Wolf12}, the conventional picture of excitation exchange between the atom and the field is clearly not valid, and the spectrum of the Rabi model and the associated dynamics has been shown to exhibit a variety of novel and significant nonclassical effects, such as initial state revivals \cite{Larson07,Casanova10}, strong atom-field entanglement \cite{Ashhab10}, and generation of photons from the vacuum through modulation of the coupling constant \cite{Liberato09} or qubit frequency \cite{Dodonov09,Beaudoin11}.

Based on the above discussion, it follows that practical realizations of the Rabi model in the various regimes of coupling strength will be important to completing our understanding of the coupling of light to matter at the most fundamental level. We suggest that quantum simulation can be a valuable tool in this respect. Indeed, an analog simulation of the Rabi model based on light transport in waveguides was realized in \cite{Crespi12}, and a way of simulating the model based on circuit QED was recently proposed in \cite{Ballester12}. Such effective realizations are important not only because they have the potential to simulate even greater coupling strengths than can be achieved at present by ``direct'' coupling between two-level systems and cavity modes, but also because they offer flexible means of controlling other model parameters, as well as convenient ways of probing the dynamics through well-defined output channels and measurements.

In this vein, we wish to propose a realization of the Rabi model with a single real atom coupled to a high finesse optical cavity mode. Two stable hyperfine ground states of the multilevel atom make up the effective two-level system, while (resonant) Raman transitions between them are induced by the cavity field and auxiliary laser fields. This allows us to simulate the Rabi model with essentially arbitrary tuning of the effective frequencies and coupling constant, so that any regime, including ultrastrong and deep-strong coupling, can be accessed and explored. Although the model could in principle be realized with a variety of (alkali) atoms, we focus on the D1 line of ${}^{87}$Rb, and consider cavity QED parameters that should be achievable with, for example, microtoroidal whispering-gallery-mode (WGM) resonators \cite{Spillane05}.

Inevitably, all such systems are subject to dissipation, and the Hamiltonian model must be expanded to an open systems treatment. We will see that our effective model is described by a master equation of the form
\begin{align}\label{eq:master:intro}
  \dot{\rho} = -i[H_R,\rho] + \kappa (2a\rho a\dagg - a\dagg a \rho - \rho a\dagg a),
\end{align}
where $\kappa$ is the cavity decay rate. On the one hand, we can realistically expect $\kappa$ to be small enough for the Hamiltonian dynamics to dominate on appreciable time scales, so that the Rabi model dynamics is prominent. On the other hand, the dissipative cavity QED setup provides us with a convenient means of observing the dynamics via the output field of the cavity. 
Furthermore, the open systems dynamics is of interest in itself, and the steady state of \eq{eq:master:intro} is known to possess some very interesting features \cite{Werlang08}. It should be noted, however, that such a master equation is incorrect for circuit QED systems genuinely in the ultrastrong coupling regime \cite{Ciuti06,Liberato09,Beaudoin11,Ridolfo12}, as it predicts unphysical results such as photon generation from the vacuum. In our case however, such a photon flux is perfectly legitimate, given the energy input from the laser fields that help to drive the Raman transitions that we use to implement our simulation of the Rabi model. Our scheme thus offers the possibility of realizing an open systems version of the Rabi model that is subject to a simple (``conventional'') interaction with the environment (i.e., cavity damping) and can hence explore certain novel features of the Rabi model beyond what might be achievable with genuine systems.

In addition, our scheme offers a generalization of the Rabi model by the addition of an extra term in the effective Hamiltonian describing a non-linear coupling between the qubit and the oscillator. This term takes the form 
\begin{align}
H_{NL} = \frac{U}{2} \sigma_z a\dagg a , 
\end{align}
where $U$ is a tunable parameter, and can be viewed as a dynamical shift of the oscillator frequency, $\omega \to \omega+U\sigma_z/2$. A coupling of this nature typically arises as an approximation to the Jaynes-Cummings model in the dispersive limit (i.e., when $|\omega-\omega_0|\gg g$), whereas for our model it can be present independent of the relative sizes of $\omega_0$, $\omega$ and $g$, and of a magnitude that may in fact be comparable to or even larger than these parameters. Although our initial focus will be on the effective realization of the Rabi model, \eq{eq:Rabi}, such that we set $U$ to zero or to a small enough value the effect of the non-linear coupling is unimportant, we will see that variation of this parameter is extremely interesting in itself and can lead to dramatic changes in the system's properties. In particular, we find two ``critical'' values of $U$ ($U = \pm 2\omega$), about which, in the deep strong coupling regime ($g/\omega \gtrsim 1$), sharp transitions occur in the quantum state of the system, marked by clear signatures in the cavity field.

Finally, we would like to comment on the connection between the model proposed here, and models simulating the collective interaction of an ensemble of two-level atoms with a single field mode. Our scheme is essentially the single atom version of the proposed realization of the Dicke model presented in \cite{Dimer07}. By showing that an effective ultrastrong coupling regime is in principle achievable on the single atom level using a ${}^{87}$Rb coupled to a high finesse optical microcavity, we also offer a specific means of realizing the many-atom Dicke model quantum phase transition as proposed in \cite{Dimer07} (which requires only strong collective coupling of $N$ atoms to a cavity mode, a much weaker requirement than the very strong single atom coupling assumed in this work). We also note the related work that has been done in \cite{Agarwal12}.

In fact, the above-mentioned experimental realization of the Dicke quantum phase transition \cite{Baumann10} was carried out using a scheme analogous to that of \cite{Dimer07}, but based upon laser-plus-cavity-mediated, resonant Raman transitions between discrete momentum states of a Bose-Einstein condensate. Notably, that particular scheme also gives rise to a nonlinear coupling term of the form $J_z a\dagg a$, where $J_z$ is the many-atom inversion operator. The Dicke model including this term has been studied theoretically in the thermodynamic limit ($N\rightarrow\infty$) in \cite{Keeling10,Bhaseen12}, where phase diagrams for the semiclassical steady state have been mapped out for parameters related primarily to the experiment of \cite{Baumann10}. The nonlinear atom-photon coupling was shown to be of fundamental importance; in particular, a new superradiant phase is possible if the effective nonlinear coupling constant is negative ($U < -2\omega$). Our scheme could also offer a flexible platform for exploring such a regime and, indeed, the sharp transitions we already see at the single atom level highlight this possibility.

\section{The Model}
\subsection{Full system}

\begin{figure}
\includegraphics{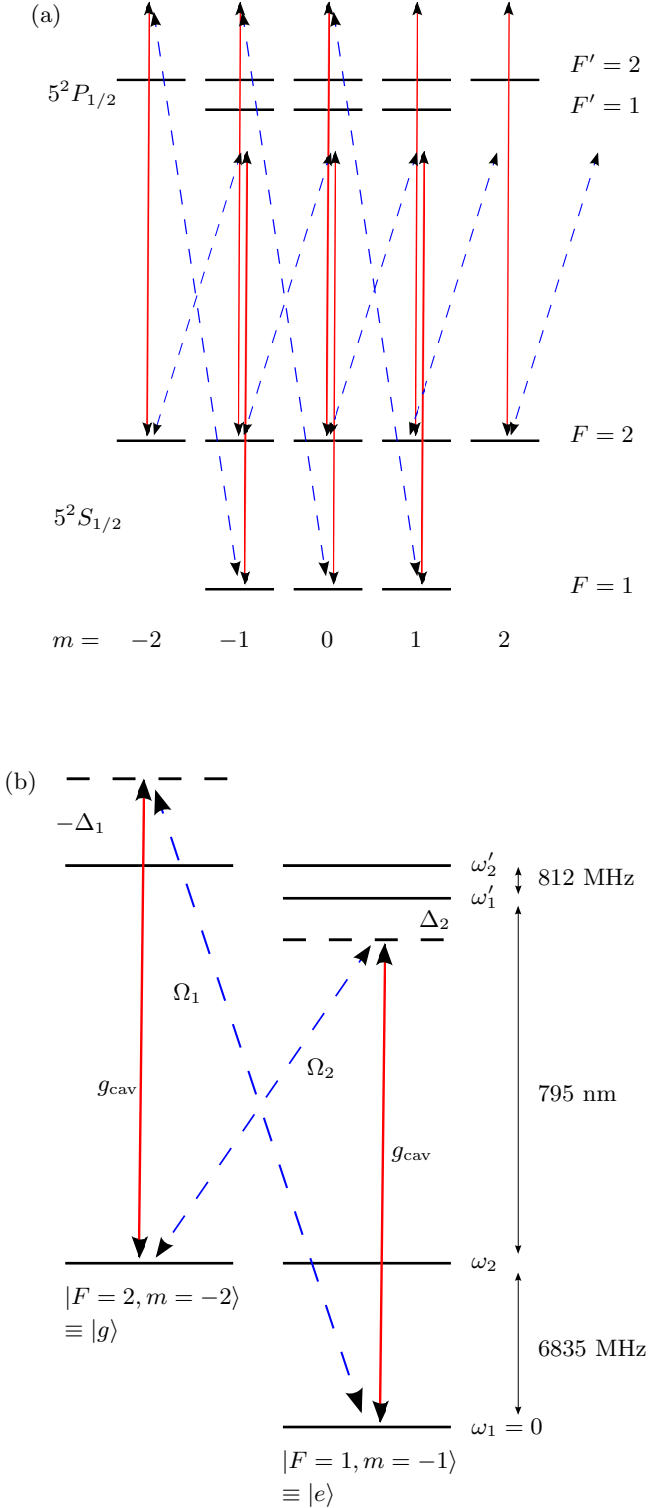}
\caption{\label{fig:levelscheme} (Color online). (a) Energy level diagram for the D1 transitions (not to scale). The two laser fields (blue dashed lines) are $\sigma_-$- and $\sigma_+$-polarized, respectively. The cavity mode (red solid lines) is $\pi$-polarized and participates in resonant (or near-resonant) Raman transitions from both the $F=1$ and $F=2$ ground states. (b) Detailed view of the leftmost part of the energy level diagram. Horizontal dashed lines indicate detunings of the fields from the excited states.}
\end{figure}

The physical configuration that we consider here employs electric dipole transitions on the D1 line of a single ${}^{87}$Rb atom. By coupling the atom simultaneously to an optical cavity mode and two laser fields, two stable hyperfine ground states -- one in the $F=1$ level and one in the $F=2$ level of the $5^2S_{1/2}$ state -- are connected through a pair of (distinct) Raman transitions. The specific scheme is illustrated in \fig{fig:levelscheme}. In particular, $\sigma_+$- and $\sigma_-$-polarized laser fields, separated in frequency by approximately twice the ground-state hyperfine splitting of $2\pi\cdot 6.835$~GHz  and each far from resonance with the $5^2S_{1/2}-5^2P_{1/2}$ transition frequency, combine with a $\pi$-polarized cavity mode to drive resonant or near-resonant Raman transitions between pairs of states, each of which consist of one state in the $F=1$ level and one in the $F=2$ level. We will eventually focus on just one pair of states, $\ket{F=2,m=-2}$ and $\ket{F=1,m=-1}$, but for the moment consider the most general model.

We introduce atomic dipole transition operators $A_{FF'}^{(p)}$ connecting level $F$ in the $5^2S_{1/2}$ state to level $F'$ in the $5^2P_{1/2}$ state with polarization $p$:
\begin{align}
  A_{FF'}^{(p)} = \sum_{m=-F}^F \braket{F,m|\mu_p|F',m+p} \ket{F,m}\bra{F',m+p} .
\end{align}
Here, $m$ labels the magnetic sublevel, $p = \{-1,0,+1\}$ denotes $\{\sigma_-,\pi,\sigma_+\}$-polarization, respectively, and $\mu_p$ is the corresponding dipole operator. The dipole matrix elements are normalized such that $\sum_{F,p}|\braket{F,m-p|\mu_p|F',m}|^2 = 1$, and numerical values for the elements can be found, for example, in \cite{Steck10}. We can then compactly write the full Hamiltonian of the system in the form (setting $\hbar = 1$)
\begin{widetext}
\begin{align}\label{eq:H:full}
  H_{D1} =& H_0 + H' , \\
  H_0 =& \omega_\text{cav} a\dagg a + \omega_2 \sum_{m=-2}^2 \ket{2,m}\bra{2,m} + \omega_1' \sum_{m=-1}^1 \ket{1',m}\bra{1',m} + \omega_2' \sum_{m=-2}^2 \ket{2',m}\bra{2',m} , \\
  H' =& \left\{ \Omega_1 \e^{i\omega_{L_1}t}\left(A_{11}^{(-1)} + A_{12}^{(-1)} + A_{21}^{(-1)} + A_{22}^{(-1)}\right) + \Omega_2 \e^{i\omega_{L_2}t}\left(A_{11}^{(+1)} + A_{12}^{(+1)} + A_{21}^{(+1)} + A_{22}^{(+1)}\right) \right. \nonumber \\
  & \left.  ~+ g_\text{cav}\left(A_{11}^{(0)} + A_{12}^{(0)} + A_{21}^{(0)} + A_{22}^{(0)}\right)a^\dagger  \right\} + {\rm H.c.}
\end{align}
\end{widetext}
Here, $\omega_\text{cav}$ is the cavity frequency, $\omega_i$ ($i=1,2$) is the frequency (energy) of the atomic $F=i$ level relative to the $\{ 5^2S_{1/2},F=1\}$ level, with primed quantities denoting the excited levels, and $\omega_{L_1}$ and $\omega_{L_2}$ are the laser frequencies. For brevity, we use the notation $\ket{i,n} = \ket{F=i,m=n}$, $\ket{i',n} = \ket{F'=i,m=n}$, while $a$ ($a^\dagger$) is the annihilation (creation) operator for the cavity mode. The laser Rabi frequencies are $\Omega_1$ and $\Omega_2$, while $g_\text{cav}$ is the atom-cavity coupling strength.
Note that the terms in $H'$ that are proportional to $\{ A_{21}^{(-1)}, A_{22}^{(-1)}, A_{11}^{(+1)}, A_{12}^{(+1)}\}$ (and their conjugates) do not participate in any resonant or near-resonant Raman transitions and are not shown in \fig{fig:levelscheme}, but these ``off-resonant'' terms can induce non-negligible shifts of the effective two-level frequency splitting that we derive below.

Including cavity (field) decay and atomic spontaneous emission at rates $\kappa$ and $\gamma$, respectively, the evolution of the system density operator, $\rho$, is given by a master equation of the form
\begin{align}\label{eq:master:full}
  \dot{\rho} = -i[H_{D1},\rho] + \kappa\mathcal{D}[a]\rho + \frac{\gamma}{2} \sum_{p,F,F'} \mathcal{D}\left[ A_{FF'}^{(p)} \right]  \rho \, ,
\end{align}
where $\mathcal{D}[O]\rho \equiv 2O\rho O\dagg - O\dagg O \rho - \rho O\dagg O$ for any operator $O$.
The spontaneous emission rate for the D1 line of ${}^{87}$Rb is $\gamma/(2\pi)= 5.7$~MHz. We are most interested in parameter regimes where the atomic excited state populations are negligible, so we will neglect atomic spontaneous emission in the effective model that we derive below. However, spontaneous emission will be included in all of our numerical simulations of the full model above and we consider its effects in \sect{sec:num_comp}.

\subsection{Reduced system}

As alluded to above, we assume very large detunings of the fields from the atomic excited states, so that these states are only ever virtually populated (assuming no initial population) and can be adiabatically eliminated to yield an effective model involving only ground states (i.e., levels in the state $5^2S_{1/2}$). To do this it is first convenient to move to a rotating frame through the unitary transformation defined by $U(t)=\exp(-iH_0t)$ with
\begin{align}
  H_0 =& (\omega_{L_2} + \tilde{\omega}_2)a\dagg a + \tilde{\omega}_2 \sum_m \ket{2,m}\bra{2,m}\nonumber \\
       &+ (\omega_{L_2}  + \tilde{\omega}_2) \sum_m \ket{1',m}\bra{1',m} \nonumber\\
       &+ \omega_{L_1} \sum_m \ket{2',m}\bra{2',m},
\end{align}
where $\tilde{\omega}_2=(\omega_{L_1}-\omega_{L_2})/2$ is a frequency close (or equal) to $\omega_2$, the ground state hyperfine splitting. We will from now on assume this transformation when referring to \eq{eq:master:full}. Defining
\begin{align}
  \Delta_1 =& \omega_2' - \omega_{L_1}  ,\\
  \Delta_2 =& \omega_1' - (\omega_{L_2} + \tilde{\omega}_2)  ,
\end{align}
the condition for the validity of the adiabatic elimination is that $|\Delta_{1,2}| \gg g_\text{cav},|\Omega_1|,|\Omega_2|,\kappa,\gamma$. Neglecting spontaneous emission and terms rotating (in the transformed frame) at frequency $\tilde{\omega}_2$, we obtain an effective Hamiltonian describing energy shifts of the ground state levels due to the various fields, and (Raman) couplings between pairs of levels, each pair consisting of one level in the $F=1$ state and one in the $F=2$ state. Focusing on the two leftmost levels as depicted in \fig{fig:levelscheme}(b), the relevant part of the effective Hamiltonian can be written
\begin{align}
  H_\text{eff} =& \frac{\omega_0}{2} \sigma_z + \omega a\dagg a + \left( g_\text{eff}^{(1)}\sigma_+ a + g_\text{eff}^{(2)}\sigma_- a + {\rm H.c.} \right)\nonumber \\
                &+ \frac{U}{2} \sigma_z a\dagg a.
\end{align}
Here we have identified $\ket{g} \equiv \ket{2,-2}$, $\ket{e} \equiv \ket{1,-1}$, introduced Pauli operators $\sigma_z = \ket{e}\bra{e} - \ket{g}\bra{g}$, $\sigma_+ = (\sigma_-)\dagg = \ket{e}\bra{g}$, and dropped constant energy terms. The parameters of the effective Hamiltonian are given by
\begin{align}
  \omega_0 =& \frac{1}{2}\frac{|\Omega_1|^2}{\Delta_1} - \frac{1}{2}\frac{|\Omega_2|^2}{\Delta_2} - \frac{1}{6}\frac{|\Omega_2|^2}{\Delta_2 + \omega_{21}'} \nonumber \\
            &+ \frac{1}{12}\frac{|\Omega_2|^2}{\Delta_2+\tilde{\omega}_2} + \frac{1}{12}\frac{|\Omega_2|^2}{\Delta_1+2\tilde{\omega}_2} + \delta \label{eq:w_0_eff} , \\
    \omega =& \delta_\text{cav} - \frac{|g_\text{cav}|^2}{2}\left(\frac{1}{3}\frac{1}{\Delta_1} + \frac{1}{4}\frac{1}{\Delta_2+\omega_{21}'} + \frac{1}{12}\frac{1}{\Delta_2}\right) , \\
       g_\text{eff}^{(1)} =& \frac{1}{2\sqrt{6}}\left(\frac{g_\text{cav}\Omega_2\conj}{\Delta_2} + \frac{g_\text{cav}\Omega_2\conj}{\Delta_2 + \omega_{21}'}\right) , \\
       g_\text{eff}^{(2)} =& \frac{1}{\sqrt{6}}\frac{g_\text{cav}\Omega_1\conj}{\Delta_1} , \\
      U = & |g_\text{cav}|^2 \left(\frac{1}{4}\frac{1}{\Delta_2+\omega_{21}'} + \frac{1}{12}\frac{1}{\Delta_2} - \frac{1}{3}\frac{1}{\Delta_1}  \right) ,\label{eq:U_eff}
\end{align}
where
\begin{align}
\delta =& \omega_2 - \tilde{\omega}_2 , 
\\
\delta_\text{cav} =& \omega_\text{cav} - (\omega_{L_2} + \tilde{\omega}_2) ,
\end{align}
and $\omega_{21}' =\omega_2'-\omega_1' = 2\pi\cdot 812$~MHz is the excited state hyperfine splitting. Note that, in fact, $\Delta_2 + \omega_{21}' = \Delta_1 + \tilde{\omega}_2$. The numerical prefactors to the various terms are products of dipole matrix elements associated with the atomic transitions \cite{Steck10}.

\subsection{Realization of the Rabi model}

We can choose $g_\text{eff}^{(1)} = g_\text{eff}^{(2)} \equiv g_\text{eff}$ (real), so that $H_\text{eff}$ reduces to the form of a generalised Rabi model,
\begin{align}\label{eq:H:eff}
  H_\text{eff} =& \frac{\omega_0}{2} \sigma_z + \omega a\dagg a + g_\text{eff}(\sigma_+  + \sigma_-)(a + a\dagg)\nonumber \\
                &+ \frac{U}{2} \sigma_z a\dagg a.
\end{align}
The term proportional to $U$ can be made small compared to the other terms (or even zero) with a judicious choice of the physical parameters. Provided this is so, $H_{\rm eff}$ achieves an essentially faithful realization of the Rabi model. Furthermore, the effective frequencies and coupling strength of the model are all determined by either level shifts or Raman transition rates, which are tunable via the laser frequencies and intensities and can therefore be chosen to be of the same magnitude.
In other words, we have a model of a two level system coupled to a single mode of the electromagnetic field where it is in principle possible to access any regime of coupling strength -- strong, ultrastrong or deep-strong coupling -- as defined by the ratios $\omega_0/\omega$ and $g_\text{eff}/\omega$. 

Of course, our realization is with an open system and, including cavity dissipation, the master equation for the evolution of the reduced system density operator is
\begin{align}\label{eq:master:eff}
  \dot{\rho}_\text{eff} = -i[H_\text{eff},\rho_\text{eff}] + \kappa\mathcal{D}[a]\rho_\text{eff}  .
\end{align}
For the full Rabi model dynamics to be observable, we clearly require that the parameters of the model exceed the cavity field decay rate $\kappa$. Hence, our proposed realization demands a strong-coupling cavity QED system, i.e., $g_\text{cav}\gg\kappa$, so that, in particular, the effective coupling strength, $g_\text{eff}$, can also exceed $\kappa$. 

Cavity QED with microtoroidal optical resonators, in which atoms close to the surface of a microtoroid couple to the evanescent fields of whispering gallery modes (see, for example, \cite{Aoki06,Dayan08}), should be capable of providing such a system. Very small mode volumes offer the prospect of electric dipole coupling strengths reaching values in the hundreds of MHz, while ultrahigh quality factors exceeding $10^8$ correspond to sub-MHz field mode decay rates at alkali atom transition frequencies \cite{Spillane05}. Below, we choose the value $g_\text{cav}/(2\pi)=200$~MHz for most of our numerical simulations of the full model, and consider field decay rates in the range $\kappa/(2\pi)=0.01-0.2$~MHz.

\subsection{Variation of Rabi model parameters}

To demonstrate how the effective parameters, Eqs.~(\ref{eq:w_0_eff})--(\ref{eq:U_eff}), can be tuned through a variation of the physical parameters, we consider a situation where the cavity coupling constant $g_\text{cav}$ is fixed. The effective nonlinear coupling $U$ can then be set by an appropriate choice of detunings $\Delta_1$ and $\Delta_2$. The linear coupling strength, $g_\text{eff}$, is tuned, for example, by varying $\Omega_1$, with $\Omega_2$ chosen so that $g_\text{eff}^{(1)}=g_\text{eff}^{(2)}$, which amounts to the condition $\Omega_2/\Delta_2 + \Omega_2/(\Delta_2 + \omega_{21}') = 2 \Omega_1/\Delta_1$. This also sets the values of the quantities $\omega_0 - \delta$ and $\omega - \delta_\text{cav}$. The effective frequencies $\omega_0$ and $\omega$ can then be tuned through variations of the parameters $\delta$ and $\delta_\text{cav}$. Since $\{ |\delta|, |\delta_\text{cav}|\} \ll \{ |\Delta_1|,|\Delta_2|,\tilde{\omega}_2\}$, the adjustment of $\omega_0$ and $\omega$ in this manner is essentially independent of the tuning of $g_\text{eff}$ and $U$. Note also that $\delta_\text{cav}$ could be adjusted simply via the cavity frequency $\omega_\text{cav}$, while an external magnetic field could also control $\omega_2$ and hence $\delta$ through the relative Zeeman shift of the levels $\ket{2,-2}$ and $\ket{1,-1}$.

\begin{figure}
  \includegraphics{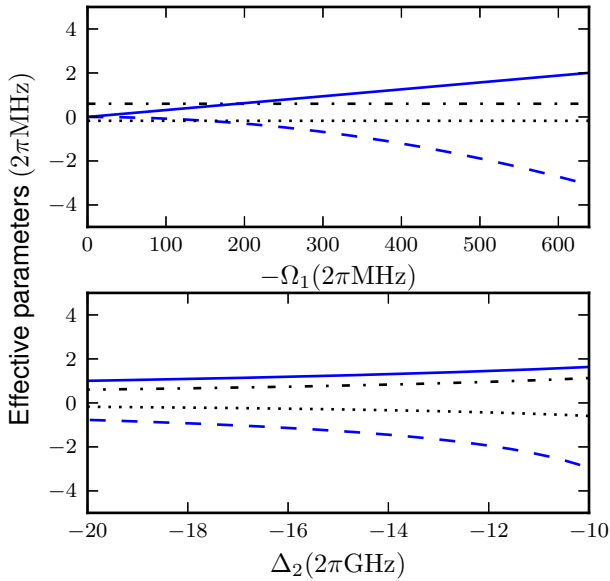}
  \caption{\label{fig:params} (Color online). Effective parameters in units of $(2\pi)\cdot$MHz: $g_\text{eff}$ (blue, solid line), $\omega_0$ (blue, dashed line), $\omega$ (black, dash-dotted line) and $U$ (black, dotted line). Top panel: $\Delta_1/(2\pi) = -26$ GHz, $\Delta_2/(2\pi) = -20$ GHz, while $\Omega_1$ is varied ($\Omega_2$ is varied simultaneously in such a way that $g_\text{eff}^{(1)}=g_\text{eff}^{(2)}$ for all $\Omega_1$). Bottom panel: $\Omega_1/(2\pi)= 320$ MHz, while $\Delta_2$ is varied, with $\Omega_2$ chosen such that $g_\text{eff}^{(1)} = g_\text{eff}^{(2)}$ for all $\Delta_2$. For both panels $g_\text{cav}/(2\pi)=200$~MHz and $\delta=\delta_\text{cav}=0$ (so $\Delta_2-\Delta_1=\omega_2-\omega_{21}'$). }
\end{figure}

Variation of the effective parameters as a function of laser Rabi frequencies and detunings is illustrated in \fig{fig:params} for $g_\text{cav}/(2\pi)=200$~MHz. For simplicity, we set $\delta = \delta_\text{cav} = 0$, but note that the curves for $\omega$ and $\omega_0$ are simply translated (uniformly) up or down for finite values of $\delta$ and $\delta_\text{cav}$. In the top panel, the detunings $\Delta_1/(2\pi) = -26$ GHz and $\Delta_2/(2\pi) = -20$ GHz are fixed (which sets a value for $U/(2\pi) = -0.18$ MHz), while the Rabi frequency $\Omega_1$ (and $\Omega_2$) is varied. 
In the bottom panel $\Omega_1/(2\pi) = -320$ MHz is fixed, while $\Delta_2$ is varied; $\Omega_2$ and $\Delta_1$ are varied correspondingly to satisfy $g_\text{eff}^{(1)} = g_\text{eff}^{(2)}$ and $\Delta_2+\omega_{21}' = \Delta_1+ \omega_2$, respectively.

\section{The Rabi model simulation\label{sec:num_comp}}

In this section we first investigate the validity of the effective model, \eq{eq:master:eff}, by comparing its numerical solution to that of \eq{eq:master:full} with suitably chosen parameters \cite{Johansson12}. We focus on the simulation of the Rabi model, \eq{eq:Rabi} (i.e., we choose small values of $U$ -- larger values will be considered in \Sect{sec:num_gz}), in the ultrastrong ($g_\text{eff}/\omega \lesssim 1$) and deep strong coupling ($g_\text{eff}/\omega > 1$) regimes, with a focus on the behavior of the mean intracavity photon number and the initial state revival probability. Having established a valid operating regime for the effective model, we then investigate briefly field-atom entanglement in the (dissipative) Rabi model. 

The open system nature of our setup makes it especially accessible to experimental investigation. We can imagine, for example, a situation in which the atom is prepared in one of the states $\ket{g}$ or $\ket{e}$, while the cavity mode is initially in the vacuum state. Then, the lasers are turned on so that the system evolves according to \eq{eq:master:full}. The cavity output can be continuously monitored by photon detectors, such that one can infer, for example, the intracavity photon number, field quadrature amplitudes, or photon correlations. The interaction can be stopped at any time by simply turning off the laser fields, after which one could also measure the state of the atom by, e.g., fluorescence detection.

\subsection{Photon number and revival probability}

\begin{table}
  %\centering
  \begin{tabular}{c}
  \begin{tabular}{c c c c c c c c}
    \multicolumn{8}{c}{Cavity and laser parameters} \\
    \hline
    Set & $g_\text{cav}$ & $\Omega_1$ & $\Omega_2$ & $\Delta_1$ & $\Delta_2$ & $\delta$ & $\delta_\text{cav}$ \\
    \hline
     I$a$ & 200 & -160 & -120 & -26000 & -20000 & 0.19 & 0.40 \\
    II$a$ & 200 & -320 & -240 & -26000 & -20000 & 0.77 & 0.40 \\
    III$a$ & 200 & -640 & -480 & -26000 & -20000 & 3.1 & 0.40 \\
    I$b$ & 200 & -100 & -65.0 & -17000 & -11000 & 0.25 & -0.034 \\
    II$b$ & 200 & -210 & -130 & -17000 & -11000 & 0.99 & -0.034 \\
    III$b$ & 200 & -420 & -260 & -17000 & -11000 & 4.0 & -0.034 \\
   \end{tabular}
  \\
  \\
  \begin{tabular}{c c c c c}
    \multicolumn{5}{c}{Effective parameters} \\
    \hline
    Set & $\omega_0$ & $\omega$ & $g_\text{eff}$ & $U$ \\
    \hline
    I$a$ & 0.0 & 1.0 & 0.5 & -0.18 \\
    II$a$ & 0.0 & 1.0 & 1.0 & -0.18 \\
    III$a$ & 0.0 & 1.0 & 2.0 & -0.18 \\
    I$b$ & 0.0 & 1.0 & 0.5 & -0.50 \\
    II$b$ & 0.0 & 1.0 & 1.0 & -0.50 \\
    III$b$ & 0.0 & 1.0 & 2.0 & -0.50 \\
  \end{tabular}
  \end{tabular}
  \caption{Parameter sets for the data of Figs.~\ref{fig:timeevol} and \ref{fig:decay}. All numbers are in units of $(2\pi)\cdot$MHz.}
  \label{tabl:params}
\end{table}

\begin{figure}[h]
  \includegraphics{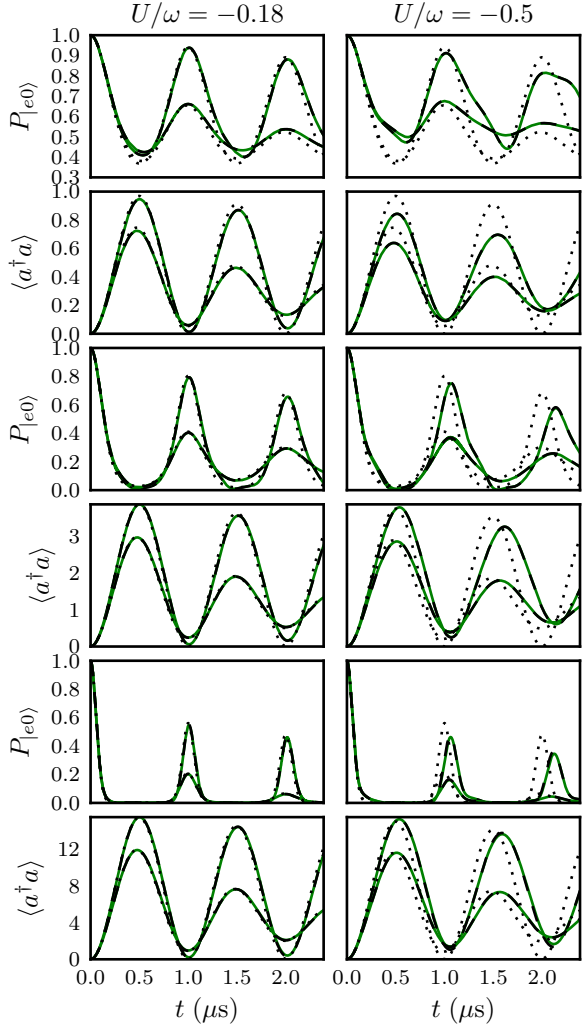}%
\caption{\label{fig:timeevol}(Color online). Time evolution of the probability of the system being in the initial state, $\ket{e0}$, and the mean intracavity photon number $\braket{a\dagg a}$. The two top rows are for the parameter sets denoted I in \tabl{tabl:params}, the two middle rows for the sets II, and the two bottom rows for the sets III. The left column corresponds to sets denoted by $a$, and the right column to $b$. The green solid lines are the solutions of \eq{eq:master:full}, and the dashed black lines are those of the effective model \eq{eq:master:eff}. The two solutions coincide perfectly to the precision of the figure. The dotted lines are for the effective model with the same parameters, but with $U$ set to zero. Each panel shows results for two different $\kappa/(2\pi) = 0.1, 0.01$ MHz -- the higher curves correspond to smaller $\kappa$.}
\end{figure}

Assuming an atom-cavity coupling strength $g_\text{cav}/(2\pi) = 200$ MHz, and choosing a selection of different laser frequencies and intensities to realize effective parameters in the ultrastrong and deep strong coupling regime, we plot in \fig{fig:timeevol} the time evolution of the probability of being in the initial state, chosen to be $\ket{e0} \equiv \ket{1,-1}\otimes\ket{0}_\text{cav}$, and the mean intracavity photon number $\braket{a\dagg a}$. The physical parameters used and the effective parameters they give rise to are shown in \tabl{tabl:params}. To consider the deviation from the Rabi model Hamiltonian due to the $U$-term in \eq{eq:H:eff}, we compare two different values: $U/(2\pi) = -0.18$ and $-0.5$~MHz. We also compare two different field decay rates, $\kappa/(2\pi)=0.1$ and $0.01$~MHz, while for the full model, \eq{eq:master:full}, the spontaneous decay rate is $\gamma/(2\pi) = 5.7$~MHz. 

Over the timescale shown in \fig{fig:timeevol}, the effects of atomic spontaneous emission are negligible, and the effective model fits the full model more or less perfectly with these parameters. We further see that, for the chosen values of $\kappa$, the Hamiltonian dynamics dominate and there are clear signatures of the deep strong coupling regime, i.e., characteristic revivals of the initial state and (periodically) large intracavity photon numbers \cite{Casanova10}. For comparison, we have also included in \fig{fig:timeevol} the evolution of \eq{eq:master:eff} with $U=0$, i.e.,  the idealized (but damped) Rabi model system, shown as dotted lines.

\begin{figure}[h]
  \includegraphics{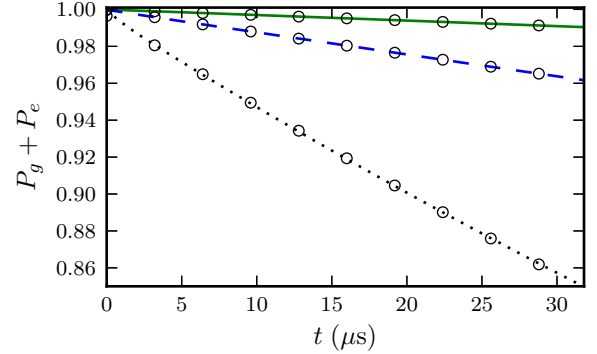}%
  \caption{\label{fig:decay}(Color online). Population decay of the $\{\ket{g},\ket{e}\}$ atomic subspace in the simulation of the full system, with $\gamma/(2\pi)=5.7$ MHz and parameter sets I$a$ (green solid line), II$a$ (blue dashed line), and III$a$ (black dotted line). Exponential fits of the form $c +a\exp(-\Gamma t)$ yield $\Gamma = 4.7\times 10^{-5}$ MHz (I$a$), $1.9\times 10^{-4}$ MHz (II$a$), and $7.8\times 10^{-4}$ MHz (III$a$). Note that $\Gamma$ scales linearly with $\Omega_i^2$ ($i=1,2$). }
\end{figure}

For longer evolution times, atomic spontaneous emission must necessarily have some effect, both within the reduced system state space $\{\ket{e},\ket{g}\}$ and by causing ``leakage'' into the other physical ground states (e.g., $\ket{2,-1}$ or $\ket{1,0}$) as well. 
To consider this effect in the analytical model, one can include the spontaneous emission term of \eq{eq:master:full} when doing the adiabatic elimination. This leads to additional dissipative terms (of Lindblad form) that connect the different physical ground states and are of  order $\bigO\left((\gamma/2) H'^2/\Delta^2\right)$, where $\Delta$ is one of $\{\Delta_{1,2},\Delta_{1,2}\pm\tilde{\omega}_2\}$. We do not present these terms here, but rather investigate effects of spontaneous emission numerically in the full model. In particular, for parameter sets I--III$a$ in \tabl{tabl:params}, we plot in \fig{fig:decay} the long-time evolution of the total probability for the atom to be in $\ket{g}$ or $\ket{e}$, given an initial system state $\ket{e0}$. The decay rates of this probability are found to be at least several orders of magnitude smaller than the rates characterizing the effective Rabi model dynamics.

While the cavity coupling constant in the above simulations is well within projected values for future microtoroidal resonators \cite{Spillane05}, it is interesting to consider still smaller values in order to explore the requirements on $g_\text{cav}$ for the simulation to be effective. We see from Eqs.~(\ref{eq:w_0_eff})--(\ref{eq:U_eff}) that a smaller $g_\text{cav}$ means the ratio of laser intensities to detunings must be larger in order to achieve the same effective coupling strength $g_\text{eff}$. This means, in general, that the adiabatic elimination is less likely to be valid and spontaneous emission will be more significant. It follows that, for any $g_\text{cav}$, there is a limit to how far we can successfully push the effective $g_\text{eff}$, or in other words the simulation scheme, into the ultrastrong regime. 

\begin{figure}[h]
  \includegraphics{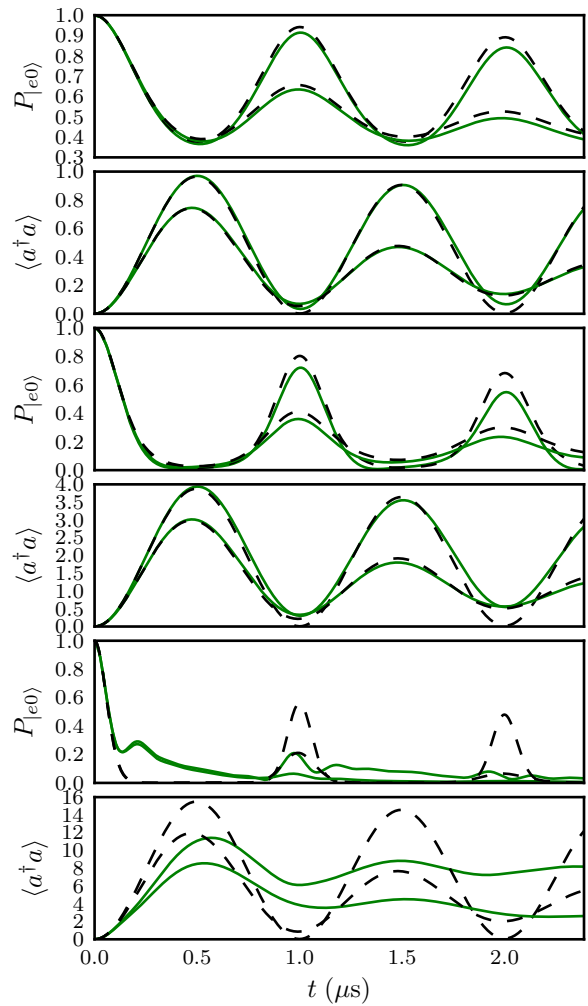}%
  \caption{\label{fig:timeevol_gc50}(Color online). Time evolution of the probability of the system being in the initial state, $\ket{e0}$, and the mean intracavity photon number $\braket{a\dagg a}$. The two top panels are for the parameter set denoted I in \tabl{tabl:params2}, the two middle panels for the set II, and the two bottom panels for the set III. The green solid lines are the solutions of \eq{eq:master:full}, and the dashed black lines are those of the effective model \eq{eq:master:eff}. Each panel shows results for two different $\kappa/(2\pi) = 0.1, 0.01$ MHz -- the higher curves correspond to smaller $\kappa$. }
\end{figure}

In \fig{fig:timeevol_gc50} we plot the time evolution of the mean photon number and the revival probability of the initial state for a physical coupling constant $g_\text{cav}/(2\pi) = 50$~MHz and parameter sets as detailed in \tabl{tabl:params2}. When comparing with the previous parameter sets given in \tabl{tabl:params}, we have essentially taken sets I--III$a$, but halved the detunings $\Delta_{1,2}$ and roughly doubled the laser intensities $\Omega_{1,2}$, so that the effective coupling strengths remain the same: $g_\text{eff}/(2\pi)=0.5$, $1.0$, and $2.0$ MHz respectively. The parameters $\delta$ and $\delta_\text{cav}$ are chosen such that $\omega_0 = 0.0$ MHz and $\omega/(2\pi) = 1.0$ MHz, as before, while the value for $U$ comes out to $U/(2\pi) = -0.04$ MHz (with this value of $U$ the solution of the effective model is indistinguishable from the case $U=0$). \fig{fig:timeevol_gc50} shows that the simulation scheme still performs very well, at least on the timescale shown, for $g_\text{eff}/\omega = 0.5$ and $g_\text{eff}/\omega = 1.0$, which are both well into the ultrastrong coupling regime. For $g_\text{eff}/\omega = 2.0$, however, it is clear that the effective model breaks down. In particular, we find a substantial loss of population from the atomic subspace $\{\ket{e},\ket{g}\}$ as a result of spontaneous emission. 

\begin{table}
  %\centering
  \begin{tabular}{c}
  \begin{tabular}{c c c c c c c c}
    \multicolumn{8}{c}{Cavity and laser parameters} \\
    \hline
    Set & $g_\text{cav}$ & $\Omega_1$ & $\Omega_2$ & $\Delta_1$ & $\Delta_2$ & $\delta$ & $\delta_\text{cav}$ \\
    \hline
    I & 50 & -390 & -230 & -16000 & -10000 & 4.4 & 0.93 \\
    II & 50 & -784 & -470 & -16000 & -10000 & 18 & 0.93 \\
    III & 50 & -1560 & -940 & -16000 & -10000 & 70 & 0.93 \\
  \end{tabular}
   \end{tabular}
  \caption{The parameter sets for the data of Figure \ref{fig:timeevol_gc50}. All numbers are in units of $(2\pi)\cdot$MHz. These sets give effective parameters $\omega_0/(2\pi) = 0$ MHz,  $\omega/(2\pi) = 1$ MHz, $U/(2\pi) = -0.04$ MHz, and $g_\text{eff}/(2\pi) = 0.5$ MHz (I), 1.0~MHz (II), 2.0~MHz (III).}
  \label{tabl:params2}
\end{table}

\subsection{Field-atom entanglement}

We conclude this section by looking at the entanglement between the two level system $\{\ket{g},\ket{e}\}$ and the field mode. Due to the realization of the Rabi model in the ultrastrong coupling regime in recent experiments in circuit QED \cite{Gunter09,Anappara09,Niemczyk10,Forn-Diaz10}, there has been interest in its application to potential quantum information technologies. In contrast to what one would expect from applying a rotating wave approximation (i.e., using \eq{eq:JC}) the ground state of the Rabi model is in general entangled \cite{Ashhab10}, and more so with stronger coupling $g_\text{eff}$, as one might intuitively expect. It is therefore interesting to consider to what extent our effective model can be used to produce entangled states between atom and field. 
The dissipative model, \eq{eq:master:eff}, does not, however, evolve the system to the ground state of $H_{\rm eff}$ (see \cite{Liberato09,Beaudoin11,Ridolfo12} for discussion and effective dissipative models that do project onto the Rabi model ground state), and the mixing due to the dissipation limits the entanglement in the system, particularly in the long-time limit.

\begin{figure}
  \includegraphics{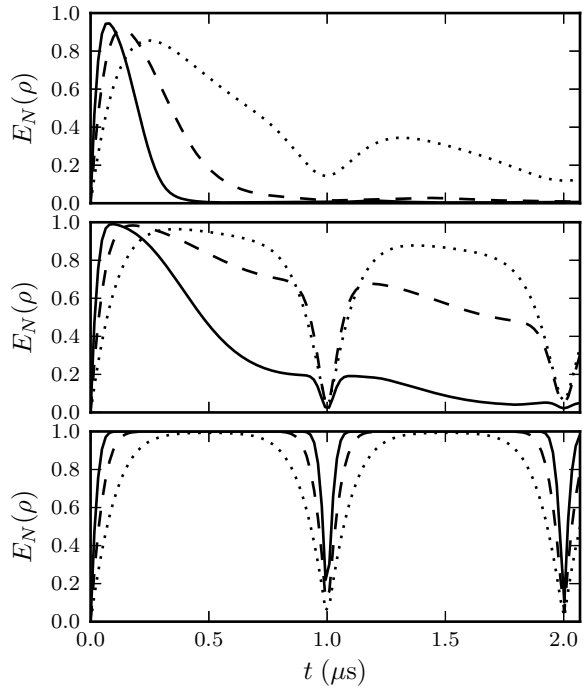}%
  \caption{\label{fig:logneg} Evolution of the logarithmic negativity, $E_N(\rho)$, for the effective model \eq{eq:master:eff} with initial state $\ket{g0}$. We consider three different coupling strengths: $g_\text{eff}/(2\pi)=2.0$ (solid lines), $1.0$ (dashed lines) and $0.5$ (dotted lines) MHz. Top panel: $\kappa/(2\pi)=0.1$ MHz, middle panel: $\kappa/(2\pi)=0.01$ MHz, bottom panel: $\kappa/(2\pi)=0.0$ MHz. Other parameters are $\omega_0=0.0$, $\omega/(2\pi)=1.0$~MHz, and $U/(2\pi) = -0.18$~MHz. }
\end{figure}

As a quantitative measure of atom-field entanglement we use the logarithmic negativity, $E_N(\rho)$ \cite{Vidal02,logneg}, which gives us an upper bound on the amount of distillable entanglement present. In \fig{fig:logneg} we plot the time evolution of $E_N(\rho)$, starting from the unentangled initial state $\ket{g0}$, for $\omega_0=0.0$, $\omega/(2\pi)=1.0$~MHz, $U/(2\pi) = -0.18$~MHz, $g_\text{eff}/(2\pi)=\{ 0.5, 1.0, 2.0\}$~MHz and $\kappa/(2\pi) = \{ 0.1, 0.01, 0.0\}$~MHz. After a short time proportional to $g_\text{eff}^{-1}$, the system evolves to a highly entangled state closely approximating the form $\ket{\psi} = \{  \ket{e}(\ket{\alpha} + \ket{-\alpha}) - \ket{g}(\ket{\alpha} -\ket{-\alpha})\}/2$, where $\ket{\alpha}$ denotes a coherent state of the field mode of amplitude $\alpha$ (for $g_\text{eff}/(2\pi)=2.0$~MHz with $\kappa/(2\pi) = 0.1$~MHz, a fidelity of 0.98 is achieved). If the lasers were to be turned off at this time and the cavity field allowed to decay, then entanglement will persist, but between the atom and the light pulse propagating in the cavity output field. Alternatively, if the lasers were turned off and the atomic state measured (by, e.g., fluorescence detection), then the cavity mode will be projected into one or other of the ``Schr\"odinger cat'' states $\ket{\psi_\pm}\propto\ket{\alpha}\pm\ket{-\alpha}$ \cite{Ballester12}.

\begin{figure}
  \includegraphics{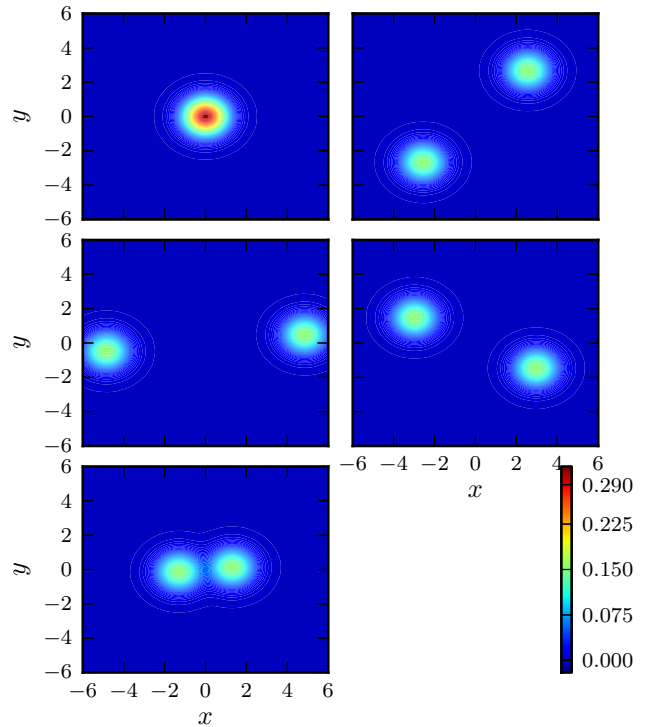}%
  \caption{\label{fig:wigner}(Color online). Contour plots of the Wigner function $W(\alpha)$, with $\alpha=(x+iy)/\sqrt{2}$, showing snapshots of the cavity field's time evolution. $g_\text{eff}/(2\pi)=2.0$ MHz, $\kappa/(2\pi) = 0.1$ Mhz, and other parameters as given in the text. Starting from the top left and traversing rows first, the times for the snapshots are $t =\{ 0, 0.25, 0.5, 0.75, 1.0\}~\mu$s. The corresponding mean photon numbers are $\braket{a\dagg a} =\{ 0, 6.8, 11.8, 5.5, 0.9\}$.}
\end{figure}

With $\kappa>0$ the logarithmic negativity eventually decays towards zero for longer evolution times at a rate that actually grows with $g_\text{eff}$. While a larger $g_\text{eff}$ gives a higher maximum entanglement at short times, it also makes the system evolution more sensitive to dissipation. This can be related to the state of the field mode $\rho_\text{cav}$, which we examine in \fig{fig:wigner} with snapshots of the time evolution of the Wigner function for the cavity field for the $\{g_\text{eff}, \kappa\} /(2\pi) = \{2.0,0.1\}$~MHz case. The Wigner function is defined by
\begin{align}\label{eq:Wigner}
  W(\alpha) = \frac{2}{\pi} \tr[D\dagg(\alpha) \rho_\text{cav} D(\alpha)(-1)^{a\dagg a}],
\end{align}
where $D(\alpha) = \exp(\alpha a\dagg - \alpha\conj a)$ is the dispacement operator and $\rho_\text{cav}$ is the reduced density operator for the cavity field. Starting from the vacuum, the Wigner function evolves into two well separated peaks, attaining a maximum separation at $t = 0.5$ $\mu$s, before the peaks nearly recombine close to the vacuum state after a full period, $t = 1.0~\mu$s. The larger the coupling $g_\text{eff}$, the greater separation of the field amplitudes corresponding to the two peaks of the Wigner function and the 
greater sensitivity of the coherence between these distinct field states to cavity decay.

\section{The non-linear atom-photon interaction \label{sec:num_gz}}

In this section we continue our investigation of the effective model, \eq{eq:master:eff}, but now examine in more detail the influence of the effective parameter $U$ on the system's behavior. The non-linear atom-photon interaction, $U \sigma_z a\dagg a/2$, can be thought of as giving an effective, dynamic shift to the cavity frequency, $\omega \to \tilde{\omega} \equiv \omega + U \sigma_z/2$, or, alternatively, to the atomic frequency, $\omega_0 \to \tilde{\omega}_0 \equiv \omega_0 + U a^\dag a$. For small $U$, the dynamics is qualitatively similar to the Rabi model, as we have seen, but as $|U|$ grows in magnitude towards the value $2\omega$ an instability develops in the system that gives rise to a sharp change in the system properties, particularly in the deep strong coupling regime. 

%that one can associate, at least qualitatively, with the potential for the effective cavity frequency $\tilde{\omega}$ to approach zero. In fact, if one considers just the generalised Hamiltonian, $H_{\rm eff}$, then for $|U|>2\omega$ the theoretical ground state is one of infinite photon number and atomic state $\ket{e}$ ($U<-2\omega$) or $\ket{g}$ ($U>2\omega$).

%Dissipation fundamentally alters this (unphysical) picture, but signatures remain in the form of dramatic changes in the system properties around the ``critical points'' $U=\pm 2\omega$ and distinct ``phases''. 

\begin{figure}
  \includegraphics{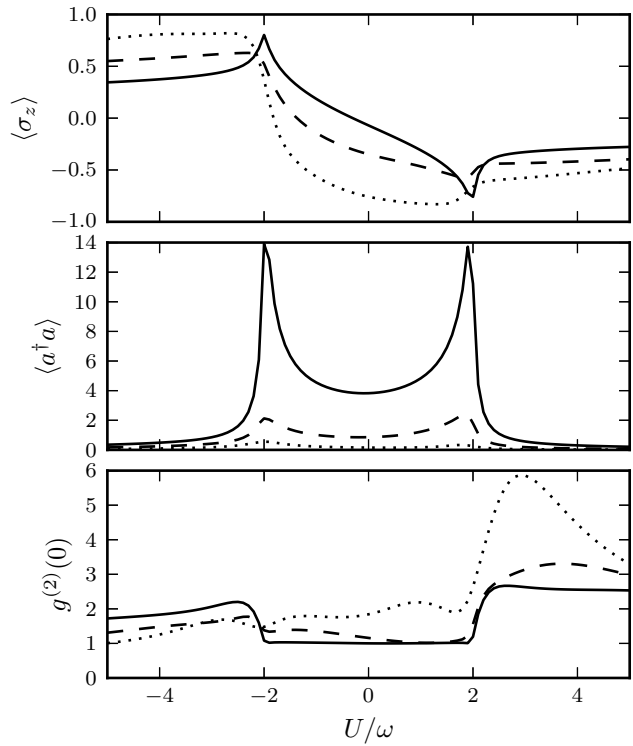}%
  \caption{\label{fig:output_N1} Steady state atomic inversion, photon number and intensity correlation function $g^{(2)}(0)$ as a function of $U$ for $g_\text{eff}/(2\pi) = 0.5$~MHz (dotted lines), $1.0$~MHz (dashed lines) and $2.0$~MHz (solid lines). Other parameters are $\omega/(2\pi) = \omega_0/(2\pi) = 1.0$~MHz and $\kappa/(2\pi)=0.2$~MHz}
\end{figure}

In \fig{fig:output_N1} we plot the steady state atomic inversion $\braket{\sigma_z}$, intracavity photon number $\braket{a\dagg a}$, and intensity correlation function $g^{(2)}(0)=\langle a^\dag a^\dag aa\rangle/\langle a^\dag a\rangle^2$ as a function of $U$. The other parameters used are $\omega/(2\pi) = \omega_0/(2\pi) = 1.0$~MHz, $\kappa/(2\pi)=0.2$~MHz, and $g_\text{eff}/(2\pi)=\{ 0.5,1.0,2.0\}$~MHz. 

\begin{figure}
  \includegraphics{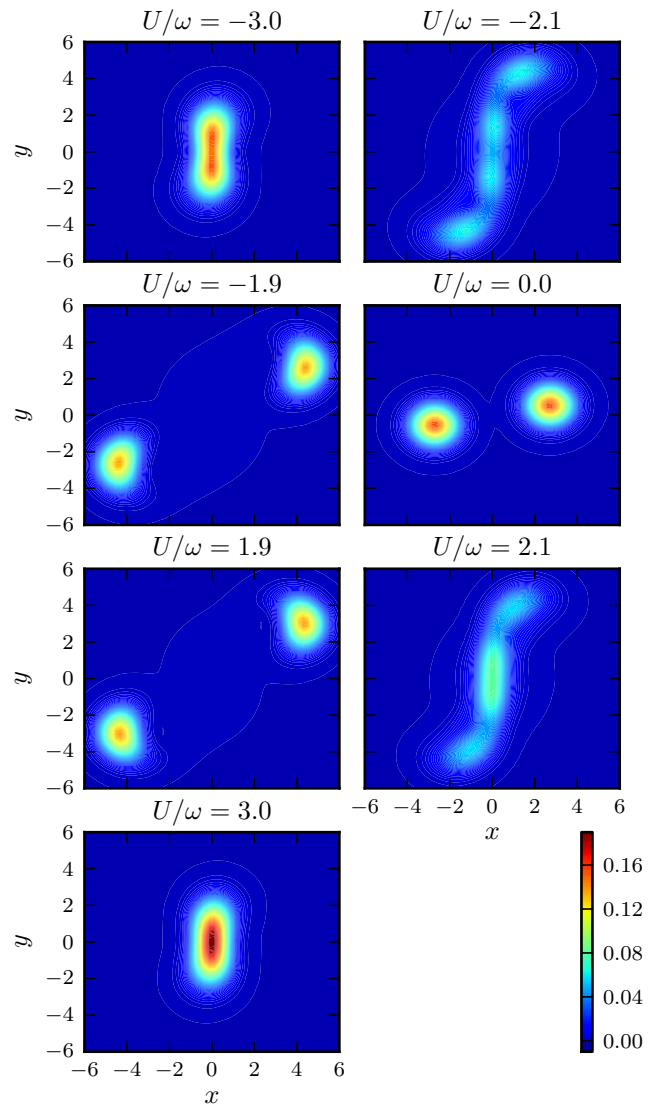}%
  \caption{\label{fig:wigner2}(Color online). Contour plots of the Wigner function $W(\alpha)$, with $\alpha=(x+iy)/\sqrt{2}$, for varying values of $U$, with $\omega_0/(2\pi)=\omega/(2\pi)=1.0$~MHz, $g_\text{eff}/(2\pi)=2.0$ MHz, and $\kappa/(2\pi) = 0.2$ MHz. }
\end{figure}

For sufficiently large $g_\text{eff}$, the behavior around $|U|=2\omega$ of each of the quantities plotted shows rapid and pronounced changes. This critical-type behavior is further exemplified by the Wigner function of the cavity field, which we plot in \fig{fig:wigner2} for $g_\text{eff}/(2\pi)=2.0$~MHz and a series of values of $U$. In particular, over a very small range of $U$ values close to $U=-2\omega$, $W(\alpha)$ changes from a weakly-split doublet aligned along the $y$-axis to a strongly-resolved doublet aligned more closely with the $x$-axis, with an intermediate (``co-existent'') phase in which $W(\alpha)$ exhibits four distinct peaks. With a finite value of $\omega_0$, the properties of the system are not symmetric in $U$ and, in particular, the Wigner function displays only a single maximum in the region $U>2\omega$. 

This phase-transition-like behaviour could be anticipated when considering the Hamiltonian dynamics given by \eq{eq:H:eff}, as there is an infinite degeneracy in the cavity mode at $\pm2\omega$, in the sense that the energy of the states $\ket{n,g/e}$ becomes independent of $n$. The degeneracy of the ground state and one or more excited states at a critical point is a signature of an equilibrium quantum phase transition in many-body systems, where one either has a level-crossing, or an ``avoided'' level-crossing that only becomes an exact degeneracy in the thermodynamic limit of an infinite number of particles \cite{Sachdev07}. Interestingly, the degeneracy of an infinite number of states at the ``critical points'' $U = \pm 2\omega$ that we consider here exists even for a single atom. Of course, the system we are considering is intrinsically open, and considering an equilibrium version is not necessarily meaningful, as this is fundamentally different from the $\kappa \to 0$ limit of the dissipative model. Indeed, for $|U|>2\omega$, the theoretical lowest energy state is one of infinite photon number and atomic state $\ket{e}$ ($U<-2\omega$) or $\ket{g}$ ($U>2\omega$). Dissipation fundamentally alters this (unphysical) picture, but signatures remain in the form of dramatic changes in the system properties.

Interestingly, a semiclassical model of the system also reveals critical behaviour at the values $U=\pm 2\omega$. This model can be derived by finding equations of motion for $\braket{\sigma_-}$, $\braket{\sigma_z}$ and $\braket{a}$ from \eq{eq:master:eff}. By assuming factorisation of operator products, $\braket{\sigma_i(a+a\dagg)} = \braket{\sigma_i}\braket{a+a\dagg}$ for $i=+,-,z$ and $\braket{\sigma_- a\dagg a} = \braket{\sigma_-}|\braket{a}|^2$, the equations form a closed set:
\begin{align}
\dot{\alpha} =& -i\left(\omega -i\kappa + \frac{U}{2} w\right)\alpha -ig(\beta + \beta\conj) \\
\dot{\beta} =& -i\left(\omega_0 + U|\alpha|^2\right)\beta + ig(\alpha + \alpha\conj) w \\
\dot{w} =& 2ig(\alpha+\alpha\conj) (\beta - \beta\conj),
\end{align}
where $\alpha \equiv \braket{a}$, $\beta \equiv \braket{\sigma_-}$ and $w \equiv \braket{\sigma_z}$. Of course, this model cannot be expected to be accurate on a single atom level where quantum fluctuations are significant, but it is relevant in the thermodynamic limit of the closely related many-atom model considered in \cite{Dimer07,Baumann10,Keeling10,Bhaseen12}. It has been studied theoretically in great detail in \cite{Keeling10,Bhaseen12} where semiclassical phase-diagrams where mapped out, with distinct superradiant phases, co-existence regions and regimes with persistent oscillations. In particular, the values $U=+2\omega$ and $U=-2\omega$ mark the change in stability of the semiclassical normal ($w = -1$, $\alpha=0$) and inverted ($w=+1$, $\alpha=0$) steady states, respectively. 

In fact, many of the features observed in the state of the cavity field (in terms of the Wigner function) in \fig{fig:wigner2} can be interpreted as manifestations on the single atom level of what become superradiant phase transitions in the thermodynamic limit: For the choice of parameters in \fig{fig:wigner2}, the semiclassical analysis predicts a superradiant phase for $U\lesssim -2\omega$ (denoted SRB in \cite{Bhaseen12}) which on the single atom level manifests itself as a weak splitting of the Wigner-function, as seen for $U/\omega = -3.0$ in \fig{fig:wigner2}. Simulations reveal that this splitting becomes more pronounced as the number of atoms in the cavity is increased, which points to a macroscopic photon number in the thermodynamic limit. Then an extremely narrow co-existence region is predicted for $U/\omega \simeq -2.0$, which we relate to the four-peaked Wigner-function for $U/\omega = -2.1$.  In the region $-2.0 \lesssim U/\omega \lesssim 2.0$ another distinct superradiant phase comes into existence (denoted SRA in \cite{Bhaseen12}), which in our case manifests itself as a more strongly split Wigner function. Finally, the region $U/\omega \gtrsim 2.0$ is a region of semiclassical persistent oscillations of $\alpha$ and $\beta$ and a value for $w$ that tends to zero as $U$ grows. This corresponds to the elongated single-peaked Wigner function for $U/\omega = 3.0$ in the figure. We defer a more detailed investigation of the connection between the semiclassical predictions, valid in the thermodynamic limit, and its comparison with quantum results for finite number of atoms, to a future work.

%In fact, for the region $|U|<2\omega$ the semiclassical treatment gives results in good quantitative agreement with the results presented above, while for $U<-2\omega$ at least qualitative agreement is found. For $U>2\omega$ it would appear that no stable semiclassical solutions exist, but we defer a detailed investigation of the semiclassical dynamics, and its comparison with quantum results for both the single- and multiple-atom system, to a future work.

\begin{figure}[t]
  \includegraphics{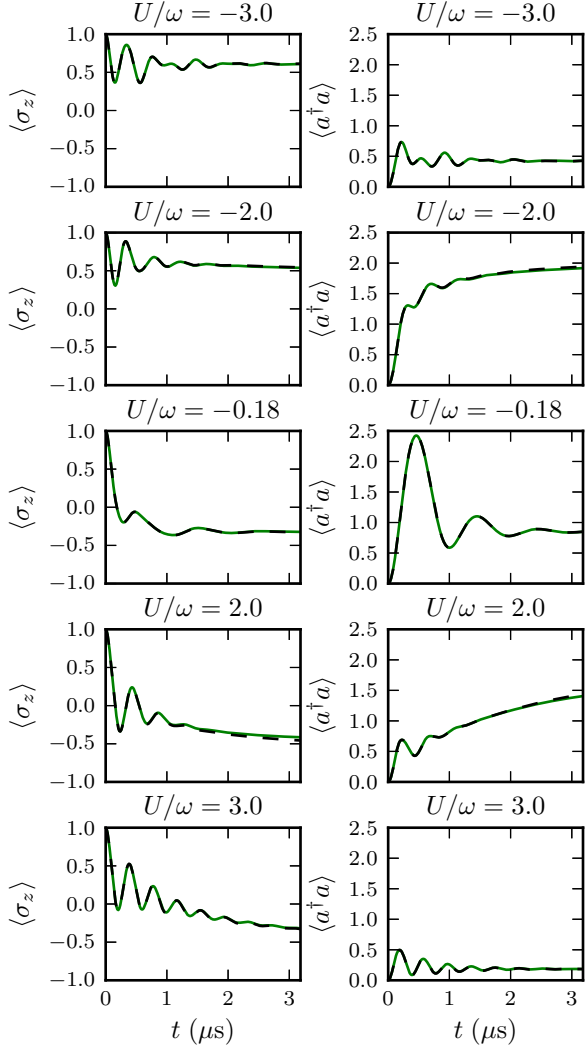}%
  \caption{\label{fig:timeevol_gz} (Color online). Time evolution of the atomic inversion, $\braket{\sigma_z}$ (solid green lines in left column), and the intracavity photon number $\braket{a\dagg a}$ (solid green lines in right column) for \eq{eq:master:full}. The dashed lines are the corresponding results for the effective model, \eq{eq:master:eff}. The solutions are nearly indistinguishable. The effective parameters are in each case set to $\omega_0/(2\pi) = \omega/(2\pi) = g_\text{eff}/(2\pi) = 1.0$ MHz, $\kappa/(2\pi)=0.2$ MHz, $\gamma/(2\pi) = 5.7$ MHz, while $U$ is varied. The corresponding cavity and laser parameters used are given in \tabl{tabl:params_gz}, with set I to V starting from the top to bottom row.}
\end{figure}

The value of $U$ also strongly influences the time-dependent behavior of the system. In \fig{fig:timeevol_gz} we plot the atomic inversion and cavity photon number as a function of time, when starting from an initial state $\ket{e,0}$, for five different values of $U$. We compare the effective model, \eq{eq:master:eff}, with the full ${}^{87}$Rb model, \eq{eq:master:full}, and find excellent agreement over the timescale shown. The physical and effective parameters used are given in \tabl{tabl:params_gz}. 
The figure shows, in particular, that characteristic timescales for oscillations and decay towards a steady state vary significantly with $U$, as one expects for a system exhibiting critical-type behavior and manifestly distinct phases associated with variations of this parameter.

\begin{table}
  \begin{tabular}{c c c c c c c c c}
    \multicolumn{8}{c}{Cavity and laser parameters} \\
    \hline
    Set & $g_\text{cav}$ & $\Omega_1$ & $\Omega_2$ & $\Delta_1$ & $\Delta_2$ & $\delta$ & $\delta_\text{cav}$ & $U$ \\
    \hline
    I & 200 & -120 & -40 & -9700 & -3700 & 1.4 & -1.9 & -3.0 \\
   II & 200 & -130 & -53 & -11000 & -4800 & 1.2 & -1.2 & -1.5 \\
  III & 200 & -320 & -240 & -26000 & -20000 & 1.8 & 0.40 & -0.18 \\
   IV & 200 & 130 & 53 & -11000 & -4800 & -0.80 & -3.2 & 1.5 \\
    V & 200 & 120 & 40 & -9700 & -3700 & -0.65 & -3.9 & 3.0 \\
  \end{tabular}
  \caption{The parameter sets referred to in \Sect{sec:num_gz}. All numbers are in units of $(2\pi)\cdot$MHz. The effective parameters realized are in each case $\omega_0/(2\pi) = \omega/(2\pi) = g_\text{eff}/(2\pi) = 1.0$ MHz, while $U$ is varied as given in the rightmost column. Notice that for set IV and V, we have used $H_\text{eff} \to -H_\text{eff}$ in \eq{eq:master:eff}.}
  \label{tabl:params_gz}
\end{table}

\section{\label{sec:conclusion}Conclusion}

We have proposed a scheme for realizing a system with dynamics described by an effective generalized Rabi model. This scheme is based on resonant Raman transitions between stable ground states of a ${}^{87}$Rb atom. By numerical solution of the master equation, including cavity decay and spontaneous emission, we have verified the feasibility of implementing this scheme, given sufficiently strong coupling of the atom to a high finesse optical cavity mode, as should be achievable with microtoroidal whispering gallery mode resonators. This scheme offers a means of performing a quantum simulation of the Rabi model with coupling strengths in both the ultrastrong ($g_\text{eff}\lesssim \omega$) and deep strong ($g_\text{eff} \gtrsim \omega$) regimes, which is interesting both from a fundamental point of view and in the context of preparing highly nonclassical states. 

Our scheme also offers a generalization of the Rabi model, with the addition of a non-linear atom-photon coupling in the Hamiltonian that can be interpreted as a dynamical shift of the cavity mode frequency. This coupling, quantified by the parameter $U$, leads to fundamentally new phases of system behavior for  magnitudes of $U$ exceeding twice the effective cavity mode frequency. Transitions between these phases are especially sharp in the deep strong coupling regime, with clear signatures in the properties of the cavity output field (i.e., in the output intensity and intensity correlation function). Extension of these results to the multiple-atom case should offer new opportunities for the study of critical phenomena in many-body cavity QED.

Finally, a possibility that we have not explored in this paper, but would like to point out, is that the effective parameters in our model could easily be made time dependent. A modulation of the effective coupling constant $g_\text{eff}$, for example, could be introduced by varying the laser Rabi frequencies $\Omega_{1,2}$ (i.e., by varying the laser intensities). In the context of the Rabi model, this has been predicted to have some fascinating consequences, such as the generation of photons from the vacuum \cite{Liberato09}. In fact, this can be viewed as being an analog of the dynamical Casimir effect, and it would be interesting to consider how our scheme could be used to simulate this elusive phenomenon.

%\appendix*
%\section{Appendix}

\begin{acknowledgments}
  ALG is grateful for the hospitality shown at the University of Auckland when the present paper was in progress. The authors thank Howard Carmichael for helpful discussions. ASP thanks Murray Barrett for discussions about potential level schemes in atomic rubidium.
\end{acknowledgments}

%\bibliography{arne}

\begin{thebibliography}{25}
\bibitem{logneg}
The logarithmic negativity is given by $E_n(\rho)=\log_2 \left\| \rho^{T_A} \right\|_1$, where $\left\| X \right\|_1={\rm Tr}\sqrt{X^\dag X}$ is the trace norm of $X$ and $\rho^{T_A}$ is the partial transpose of $\rho$ with respect to subsystem $A$.
\end{thebibliography}

\end{document}